# On the potential for reconstructing the moisture field using commercial microwave network data

Noam David[1], Omry Sendik[2], Hagit Messer[3], Huaizhu Oliver Gao[1]*, Yoav Rubin[4], Dorita Rostkier-Edelstein[5], and Pinhas Alpert[4]

*Corresponding Author: Prof. H. Oliver Gao

Email: hg55@cornell.edu

Phone: +1607-2548334.

The School of Civil and Environmental Engineering, Cornell University, Ithaca, NY 14853.

---

[1] The School of Civil and Environmental Engineering, Cornell University, Ithaca, NY, USA.
[2] The School of Computer Science, Tel Aviv University, Tel Aviv, Israel.
[3] The School of Electrical Engineering, Tel Aviv University, Tel Aviv, Israel.
[4] The School of Geosciences, department of Geophysics, Tel Aviv University, Tel Aviv, Israel.
[5] The Environmental Sciences Division, Israel Institute for Biological Research, Ness-Ziona, Israel.

**ABSTRACT**

The atmospheric greenhouse effect and the hydrological cycle of Earth are key components enabling the planet to support life. Water vapor is a central element in both of these, accounting for approximately half of the present day greenhouse effect, and comprising the most important gaseous source of atmospheric infrared opacity. Specifically, it functions as the fuel for the development of convective storm clouds. This parameter, however, is considered one of the least studied due to the limitations of conventional monitoring instruments. The current predominate monitoring tools are humidity gauges and satellites which suffer from a lack of spatial representativeness and difficulties in measuring at ground level altitudes, respectively. This study demonstrates for the first time the potential to reconstruct the 2-Dimensional humidity field using commercial microwave links which form the infrastructure for data transmission in cellular networks. Water vapor attenuates the waves transmitted by the system and thus these microwave links can potentially form as a virtual network for monitoring the humidity field. The results show a correlation of between 0.6 and 0.92 with root mean square differences ranging from 1.9 to 4.15 $gr/m^3$ between conventional humidity gauges and the humidity estimates calculated for the same points in space by the proposed technology. The results obtained are the first to point out the improved performance of humidity measurements when using data from multiple microwave links. These results indicate the tremendous potential of this novel approach for improving the initialization of meteorological forecasting models thus potentially improving the ability to cope with the dangers associated with extreme weather.

# 1 INTRODUCTION

Atmospheric humidity is one of the dominant greenhouse gases and constitutes the most important gaseous source of atmospheric infrared opacity (AGU, 1995; Held and Soden, 2000; Schimdt et al., 2010, Müller et al., 2015). Water vapor is advected in the atmosphere, redistributing energy through evaporation and recondensation. As such, it plays a fundamental role in the atmospheric vertical stability and, particularly, in the evolution of convective storm systems (Trenberth, 1999, Fabry, 2006; Weckwerth et al., 2008) along with their associated increased runoff and risk of flooding (Morin et al., 2009; David et al., 2013a).

However, current instruments for moisture monitoring, including predominantly in situ sensors and remote sensing onboard satellites or ground based, suffer from several limitations - such as poor spatial representativeness and resolution (WMO, 2008; Fabry, 2006), or limitations for measuring surface level values (Sherwood et al., 2010) - leading to poor characterization of the Atmospheric Boundary Layer (ABL) moisture field. Consequently, cardinal questions such as the magnitude of small scale fluctuation in humidity in the ABL, and its impact on convective initiation are still left unanswered (Weckwerth, 2000; Sherwood et al., 2010). Numerical weather forecasts provide key information to decision support systems that depend on atmospheric conditions. The accuracy of these forecasts depends, among other factors, on the precision of the initial conditions that are generated by assimilation of observations into the model. Specifically, atmospheric moisture plays a central role in the initialization of these models (Ducrocq et al., 2002).

Thus, developing techniques for measuring ABL moisture at a high tempo-spatial resolution, may open a window to understanding these issues, and provide decisive improvements in the ability to forecast convective storms, and provide early warning of the expected danger associated with it.

Commercial Microwave Links (MWLs) are affected by a range of atmospheric hydrometeors, primarily rainfall, and specifically by the water vapor density. These networks of links comprise the infrastructure for data transmission between cellular base stations. Since commercial MWLs are widely deployed by communication suppliers at ground level altitudes, they have the potential to provide extensive spatial observations of the moisture field through the measurement of multiple link systems.

Figure 1 shows a communication mast, the microwave antennas appear as white circles.

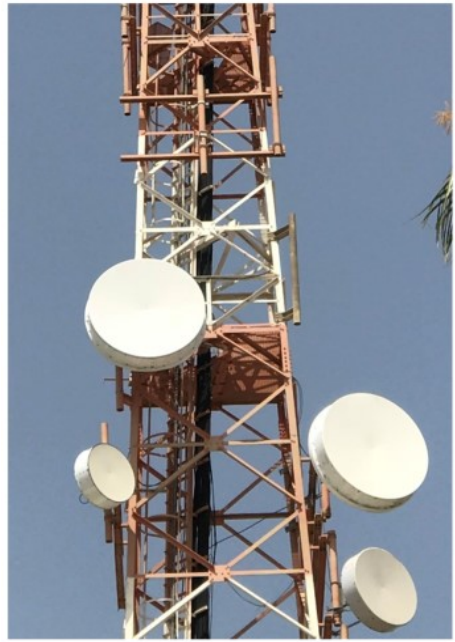

**Figure 1**: A microwave communication mast.

During the past decade it has been shown that these systems are particularly suitable for rainfall monitoring (Messer et al., 2006; Leijnse et al., 2007; Chwala et al., 2012; Rayitsfeld et al., 2012; Uijlenhoetet al., 2018) and mapping (e.g. Zinevich et al., 2009; Overeem et al., 2013). The potential for monitoring phenomena other than rain, including fog (David et al., 2013b; 2015; David and Gao, 2018), dew (Harel et al., 2015) and air pollution (David and Gao, 2016) has also been presented. Other works, point to possible applications as using commercial microwave links in urban drainage modeling (Fencl et al., 2013) and for early warning against flash flooding in Africa (Hoedjes et al., 2014). Specifically, feasibility studies conducted recently demonstrated the ability of a single commercial microwave link to acquire humidity observations (David et al., 2009; 2011).

In this study we show the potential of this opportunistic sensor network to provide extensive observations of the moisture field. The first part of the paper demonstrates a 2D theoretical simulation reveling the current and future potential of a given microwave link system in inferring the minimal detectable water vapor density at any given geographic location over a country-wide scale. The second part demonstrates the ability of retrieving improved humidity estimates in the ABL utilizing numerous MWLs deployed in different locations.

## 2 THEORY

### 2.1 Water vapor induced attenuation

MWLs operate at frequencies of tens of GHz, and as such, are influenced by atmospheric water vapor which causes signal attenuation. The relation between specific gaseous attenuation - $\gamma$ (dB/km), and water vapor density – $\rho$ (gr/m$^3$) is described by the following equation (Rec. ITU-R P.676-6, 2005; David et al., 2009):

$$\gamma = A_W + A_0 = 0.1820 \, fN''(f, P, T, \rho) \quad (dB/km) \qquad (1)$$

Where:

$A_w$ - Water vapor attenuation (dB/km)

$A_o$ - Dry air attenuation (dB/km)

$f$ - Link frequency (GHz)

$N''$ - Imaginary part of the refractivity (N-units)

P - Dry air pressure (hPa)

T - Temperature (k)

$\rho$ - Water vapor density (gr/m$^3$)

A full description of the functions comprising the expression can be found in the literature (Rec. ITU-R P.676-6, 2005). The rigorous development of the physical relation between the signal attenuation and water vapor density is given in (David et al., 2011).

Figure 2, generated based on Equation (1), presents the humidity induced attenuation for typical links of 1 km length (Figure 2(a)), and 4 kms (Figure 2(b)) as a function of system operating frequency. The dashed horizontal line represents a typical magnitude resolution of commercial microwave systems (0.1 dB) and specifically, that of the system used in this current work. The most common frequency band in operation today is between 6 and 40 GHz, and is indicated by the yellow bar. Due to the need for transferring data at higher rates, there is a trend of moving to and combining links operating in the so called e band that provides a possible cost effective solution to these increased technical demands (e.g. Csurgai-Horváth and Bitó, 2010; David et al., 2015). This band lies between 60 and 90 GHz, and is indicated by a grey bar.

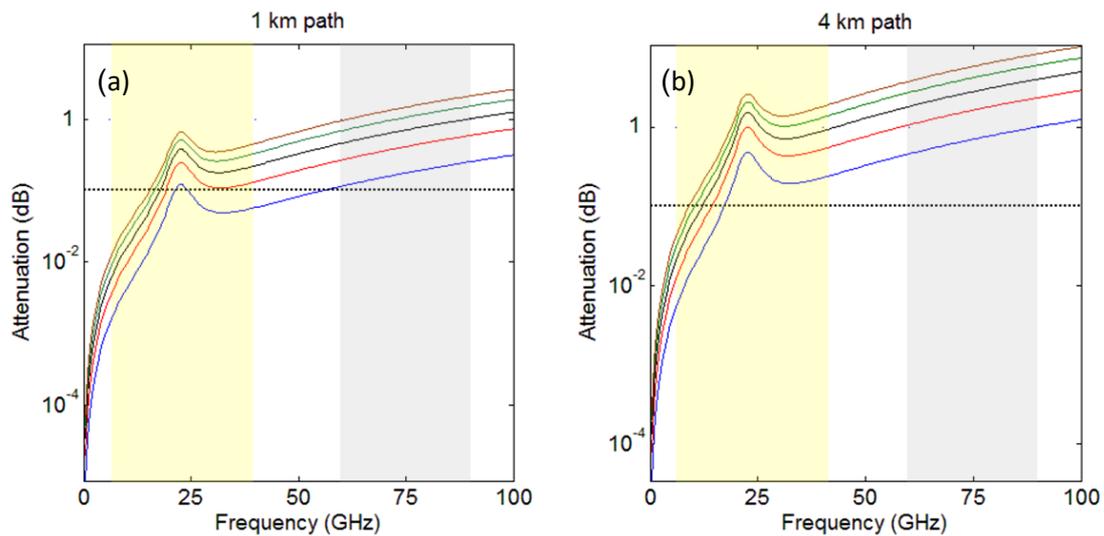

**Figure 2:** Transmission loss due to water vapor. The calculation was made for links of 1 km length (a) and 4 km length (b). The different curves represent different typical humidity at densities of 5 gr/m$^3$ (blue), 10 gr/m$^3$ (red), 15 gr/m$^3$ (black), 20 gr/m$^3$ (green) and 25 gr/m$^3$ (brown). The horizontal line represents the magnitude resolution of the microwave system used in this work (0.1 dB).

The water vapor absorption line at 22.235 GHz (Van Vleck, 1947, Chwala et al., 2014) is noticeable. We also note that at the higher end of the e band, humidity induced attenuation is greater than the induced attenuation around this absorption line, and hence the future potential of using this frequency range for humidity measurements. The magnitude resolution is constant per link (0.1 dB), thus the effective sensitivity per distance unit is higher for longer links as can be seen from the distance of the attenuation curves from the dashed horizontal line (representing magnitude resolution) in Figures 2(a) and 2(b), respectively.

*2.2 Theoretical sensitivity test*

For a given link frequency and magnitude resolution, the effective sensitivity of the link per unit of distance is greater the longer the link is (Fig. 2). The simulation tool developed in this part of the research is based on this principle. Using Eq. (1) the minimal density of water vapor that can be detected by a link network deployed over a given area is calculated. The algorithm receives the real locations of a set of MWLs, as they are spread over an area in the field. Then, the simulation covers the entirety of the tested area with ellipses simulating patches of water vapor. If a certain vapor patch simultaneously covers several links, the algorithm selects the link that has the greatest intersection with the patch. In the next stage, based on the maximal length chosen, the algorithm calculates the minimal water vapor density that can be sensed at that certain location. This, by dividing measurement resolution ($\Delta Q$) by the maximal intersection length found (L). That is, the minimal detectable density of water vapor, is one that induces attenuation greater than $\Delta Q$. If only one link is under the patch – the calculation is carried out using the length of that link. If no links are present in a given area, it is left noted as an area where no water vapor can be detected.

The tool can carry out the calculation for different link locations and densities, varying frequencies, different magnitude resolutions, and various ellipse eccentricities and sizes, according to the user's definitions.

## 3. MATERIALS AND METHODS: USING COMMERCIAL MWLs AS A MULTI-SENSOR ARRAY

### *3.1 Link Calibration*

In order to estimate the humidity induced attenuation from each link, a reference RSL (Received Signal Level) must be set. The assumption is that when the RSL value is lower than the reference value – the additional attenuation is caused by atmospheric humidity. The typical water vapor density in the observed area of this research is between ~3 and ~25 gr/m$^3$ (Source: Israeli Meteorological Service data base) and induces attenuation of between ~0.1 and ~0.6 dB, respectively, across a propagation path of 1 km at a frequency of 22 GHz. Thus, attenuation values that lie outside of this range are typically not caused only by this phenomenon. In such possible cases, additional meteorological data can be used to ascertain the cause of the interference (e.g. rainfall). The calibration was carried out based on a time period of two weeks prior to the period where actual measurements were going to be taken. For the calibration period, the humidity in the observed area can be estimated by using the estimates of conventional monitoring tools such as humidity gauges (as done in this study) or model predictions. Given a set of N humidity estimates: $\rho_1, \rho_2,..., \rho_N$. We define $\rho_{Med}$ to be the median of the set of measurements:

$$\rho_{Med} = \text{Median}[\rho_1, _2\rho,...\rho_N] \quad (gr/m^3) \quad (2)$$

Let $\gamma_{Med}(T, P, f, \rho_{Med})$ calculated according to Eq (1), be the attenuation induced by the typical dry air and the humidity $\rho_{Med}$.

Given a set of M commercial microwave links deployed in the observed area with lengths: $L_1, L_2,…,L_M$. For the *i*th link taking *k* measurements, the reference RSL is defined as:

$$RSL_{i\_ref}= \text{Median} [RSL_{i1}, RSL_{i2}, …, RSL_{ik}]+ \gamma_{Med} \quad (dBm) \quad (3)$$

### 3.2 Estimating humidity induced attenuation for a single link

Using Eq (1) and with calibration of the link, water vapor density can be derived numerically given the humidity induced attenuation (David et al., 2009):

$$\hat{\gamma} = 0.1820 \, fN''(f, P, T, \hat{\rho}) \quad (dB/km) \quad (4)$$

where $\hat{\gamma}$ is calculated by subtracting $RSL_{i\_ref}$ from the RSL measurements of the *i*th link and, in return, $\hat{\rho}$ is the estimated water vapor density.

For given temperature and pressure values, there is a maximum physical quantity of water vapor that air can carry. In cases where the humidity calculated based on the microwave link measurements was found to be higher than the maximum expected physical value (as a result of different error factors in the link attenuation measurements, such as, dew causing wetness on the antennas, etc., as discussed in the Summary section), the algorithm places the maximum physical value for humidity in its place. Equation (5) shows the manner in which the absolute humidity was calculated given the temperature and relative humidity measurements from the different stations. The absolute maximum humidity was calculated by substituting 100% for the value of the relative humidity, and taking the maximum temperature measured in the observed area during that time period (David et al., 2009):

$$\rho = 1324.45 \times \frac{RH}{100\%} \times \frac{\exp\left(\frac{17.67T}{T+243.5}\right)}{T+273.15} \quad (5)$$

For additional reading regarding uncertainty in humidity measurements using commercial microwave links and specialized humidity, pressure and temperature sensors, the reader is referred to the literature (David et al., 2009; 2011).

### 3.3 Estimating humidity in two dimensional space using multiple links

In order to calculate the estimated humidity at a certain point in space based on the contribution of the entire link system, we used Shepard inverse distance weighted interpolation (Shepard, 1968; Goldshtein et al., 2009). We converted each link path to three points located at the two ends of the propagation path and at its center. The humidity estimate at a certain point in space was then derived based on the contributions of all of the points (links) weighted by the weighting factor $W_j$ for any given point in space *(x, y)* as follows:

$$W_j(x,y) = \begin{cases} \dfrac{(1-d_j/R)^2}{(d_j/R)^2} & \dfrac{d_j}{R} \leq 1 \\ 0 & \dfrac{d_j}{R} > 1 \end{cases} \qquad (6)$$

Where:

$d_j$ – The distance between the grid point *(x, y)* and the *j-th* observation (km)

R – The radius of influence (km) defined to be the distance at which $W_j$ reduces to zero.

$W_j$ is effectively an inverse square distance weighting function. Thus, the closer a microwave measurement based observation is to the required spatial point, the larger its weight.

Finally, the humidity estimate, $\tilde{\rho}$, at the selected point in space *(x, y)* as calculated using all 3M $\hat{\rho}_j$ observations is:

$$\tilde{\rho}(x,y) = \frac{\sum_{j=1}^{3M} w_j \hat{\rho}_j}{\sum_{j=1}^{3M} w_j} \qquad (gr/m^3) \qquad (7)$$

## 4 RESULTS AND DISCUSSION

We present the research results of the theoretical sensitivity simulation and the real data measurements using multiple links as an already existing sensor array for estimating the humidity field in space.

### *4.1 Theoretical 2D simulation*

In this part of the research we examined the system's potential for 2 dimensional mapping of atmospheric humidity on a country wide scale.

The minimal density of water vapor that could be sensed by the system at every geographical location across Israel, which was used as the test area, was calculated using the 2D simulation tool described in section 2.2 Simulated circles with a radius of ~5 kilometers representing patches of water vapor were introduced into the area, until they covered the entire observed region.

The simulation was run for operating frequencies of 22 and 86 GHz which represent, respectively, a typical operating frequency in current microwave networks, and an operating frequency which is growing more common these days (Ericsson, 2015). Since the current systems of the cellular operators in the region do not operate at the e band yet, the calculations were based on the locations and configuration of real MWLs operating in the range between 21-23 GHz (i.e. near the absorption frequency of water vapor at 22.2 GHz).

Fig. 3a shows the deployment of the links and Fig. 3b the simulation run at 22 GHz based on that deployment map. The system comprises ~1500 links, with lengths between 60 m and 10 km. Figure 2b shows the minimal density of water vapor that can be sensed by the system, where the dynamic sensitivity range is given by the scale to the right of the map.

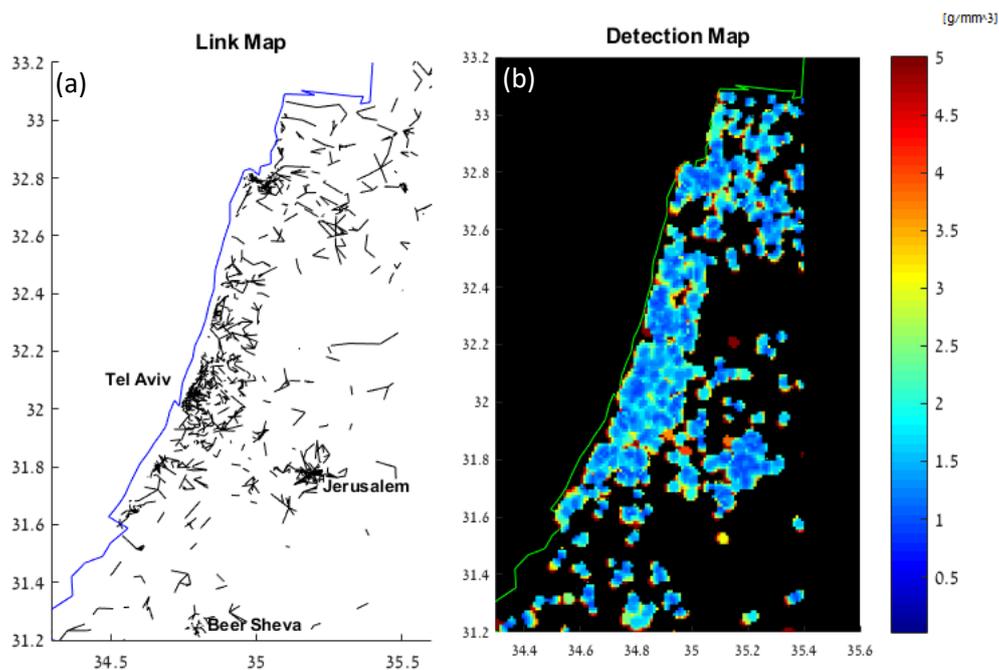

**Figure 3:** The minimal water vapor density that can be sensed at a frequency of 22 GHz. Fig. 3a shows the deployment of the MWLs, and Fig. 3b the theoretical sensitivity map calculated based on 2D simulation tool. The red color found on the upper edge of the scale bar indicates humidity values of 5 gr/m$^3$ and above.

In the next stage we compared the theoretical capability at 22 GHz with the expected one at 86 GHz. We note that typically, at higher operating frequencies, link lengths are shorter (Ericsson, 2015). Thus, MWLs that operate in the e band are not expected to be longer than a few single kilometers, and therefore, the simulation in this case was based on the same link deployment map as before, but with links longer than 4 kilometers removed. In order to create a balanced comparison, this limitation was applied to the simulation run for the 22 GHz frequency as well. Fig. 4a presents the deployment of the links used while running this simulation, Fig. 4b shows the sensitivity map at 22 GHz. Use of higher frequencies may potentially allow for monitoring with higher sensitivity, as shown in Fig 4c.

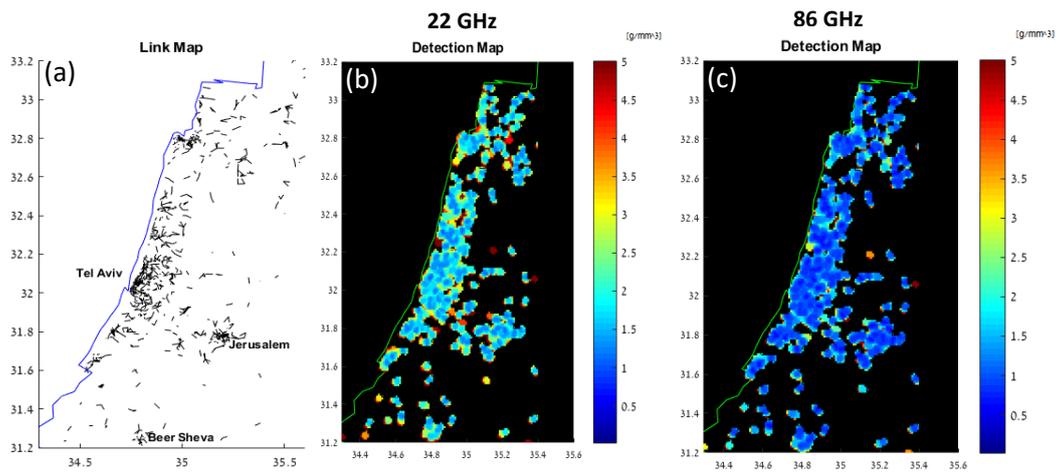

**Figure 4:** The minimal density of water vapor that can be sensed at 22 and 86 GHz. Fig.3a shows the microwave link deployment, Fig.3b the sensitivity map calculated for an operating frequency of 22 GHz, and Fig. 3c the sensitivity map at 86 GHz. The red color found on the upper edge of the scale bar indicates humidity values of 5 gr/m$^3$ and above.

## *4.2 Feasibility test: Reconstruction of the moisture field using real RSL measurements*

We concentrated on the environs of the Haifa bay in Northern Israel, where several meteorological stations are deployed in an area which is also widely covered by a network of MWLs. The microwave system used for the measurements comprises of links which provide a single instant RSL sample, once daily, at 00:00 UTC. The combined measurements from all of the links were compared to the humidity measurement taken by each of the humidity gauges at the same hour. Rainfall is a dominant factor that causes microwave attenuation. Therefore, in this part of the experiment, we used the Israeli Meteorological Service's rain data from the test site to rule out the occurrence of this phenomenon while taking moisture measurements.

Fig. 5 presents the observed area where a microwave network of 42 links operating at frequencies between 21-23 GHz as well as 10 humidity gauges are deployed. Link calibration was done based on the humidity measurements taken at the Ein Hashofet meteorological station (indicated by orange dot). The link propagation paths are drawn as lines.

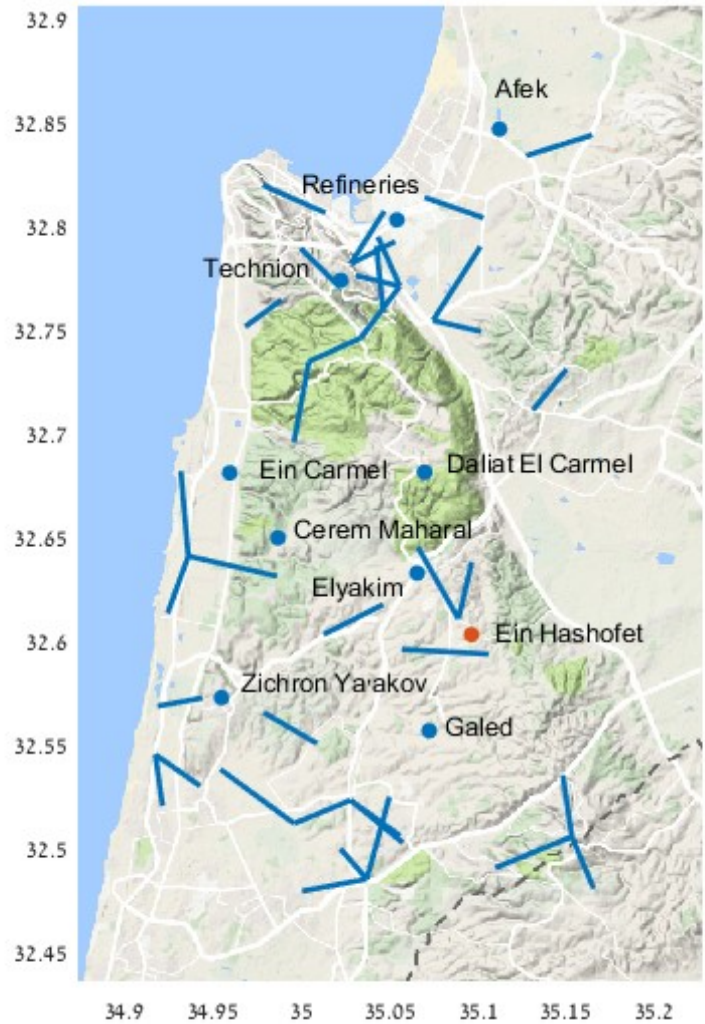

**Figure 5: The test area. The MWLs deployed in the area are at elevations of between 10 and 546 meters ASL, and with lengths between 2.1 and 4.9 kilometers. The humidity gauges, depicted as dots, are located at elevations of between 5 and 390 meters ASL.**

Figure 6 demonstrates the measurement conversion method for a single microwave link (the link that appears horizontal in the map (Figure 5) and is nearest the Ein Hashofet station). For the purpose of illustration, we deliberately chose a MWL that included outlier measurements for demonstrating the algorithm's operation. Figure 6a presents the raw RSL measurements from that link during the month of October 2013, where the dashed horizontal line represents the reference value calculated (using Equation 3) such that the induced attenuation calculated, is in fact the difference between the reference line (dashed) and the RSL measurement for each day. Figure 6b shows the humidity values as calculated from the link measurement according to Equation (4) (the continuous red line). Humidity measurements from the Ein Hashofet station are also shown (in black) for comparison, where the correlation between the two graphs was found to be 0.83. The part of the red line shown dashed, indicates estimates which deviated from physical values and were taken as outliers. These estimates were replaced by the algorithm with the maximum humidity values calculated for that time period.

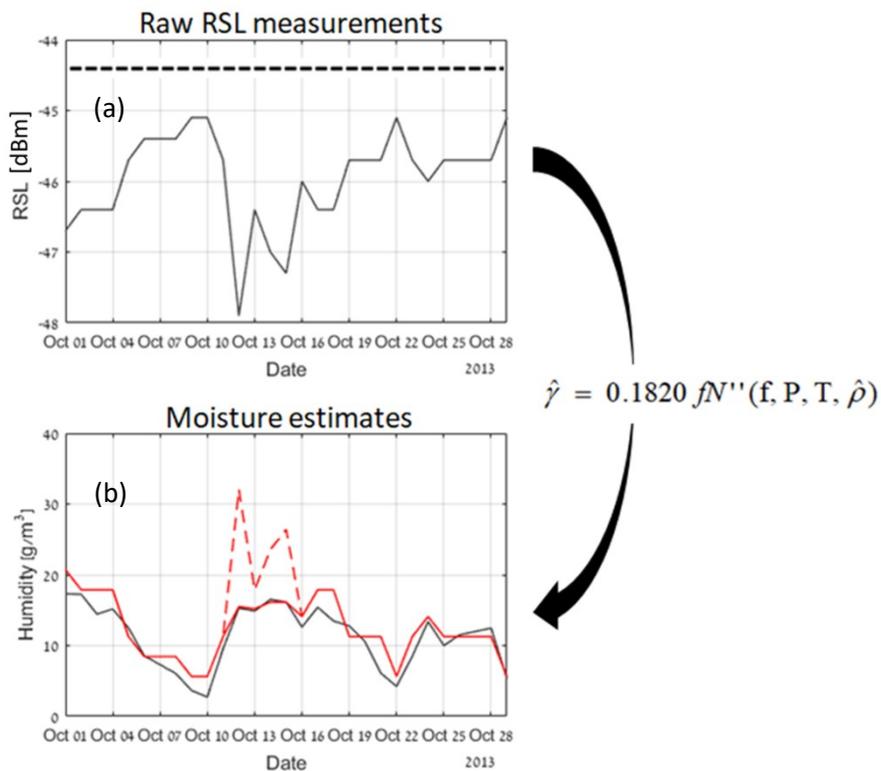

**Figure 6:** **Measurements of a single microwave link. Figure 6a shows the raw RSL measurements from the link, and Figure 6b the humidity estimates calculated from those values (continuous red line) compared with the humidity measurements from the Ein Hashofet station (in black). The dashed line indicates moisture estimates which were taken as outliers.**

Figure 7 presents the absolute humidity measurements as taken based on the MWLs (in black) compared with the humidity gauge measurements from the different meteorological stations (in red). The graphs represent an absolute humidity measurement taken once a day by each of the techniques during a month period (October 2013). The reference RSL was calculated based on 2 week measurements taken by each link during the month of September 2013. The radius of influence was taken to be R=40 km, i.e. long enough to include all MWLs deployed across the test site. The Root Mean Square Difference (RMSD, David et al., (2009)) and the Pearson correlation coefficient quantifying the fit between the different measurement techniques are located at the top of each panel. The correlation and RMSD values between the measurements of all of the links and the humidity gauge measurements in the test area were found to be between 0.6 and 0.92, and between 1.9 and 4.15 gr/m$^3$.

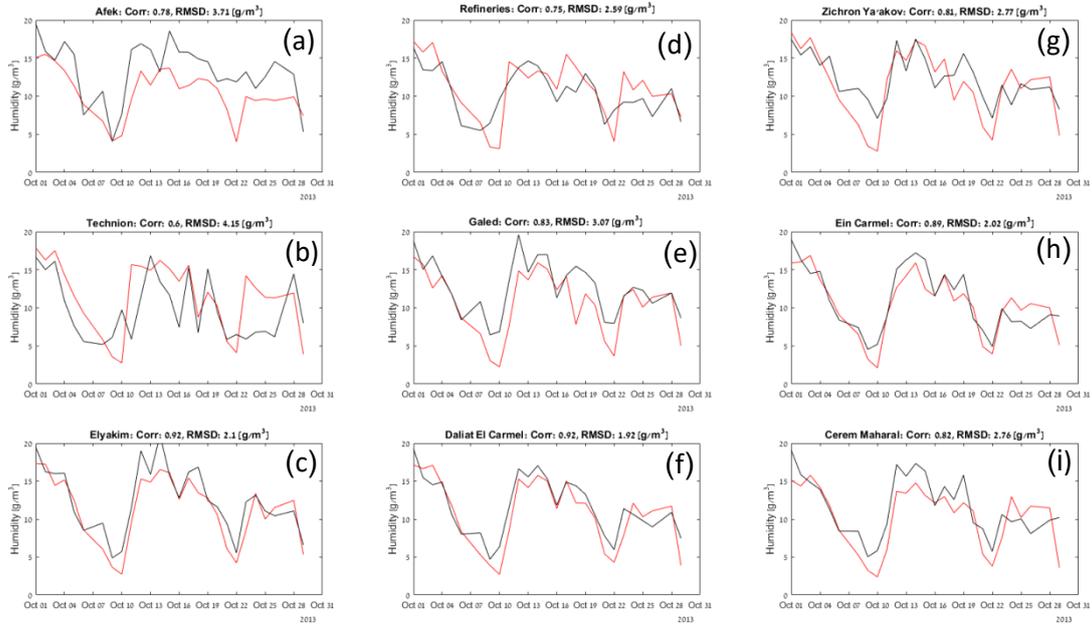

**Figure 7**: Microwave measurements compared to humidity gauge measurements. Each of the panels (a) through (i) presents the measurements taken by the different humidity gauges deployed in the test area (in red) compared with the humidity estimate derived from the weighted combination of the measurements of all of the links at that spatial point (in black).

*4.3 Analysis*

We tested the moisture estimation ability at points where humidity gauges were located, using the method proposed here, which is based on a large number of links (Equation 7) when compared to humidity estimates at the same points using a single link each time. Figure 8 shows the correlation between humidity estimates derived from weighting all of the links at a certain point where a humidity gauge was located compared to the correlation of the humidity measurements taken by each separate link from the link group, and the humidity gauge. Additionally, the horizontal axis presents the distance between the center of the link and the point where the humidity gauge is located.

It can be seen that the correlations based on weighting all of the links (the horizontal line) are significantly higher (excepting a small number of cases) than the correlations between the measurements of each single link (circles) and the station measurements.

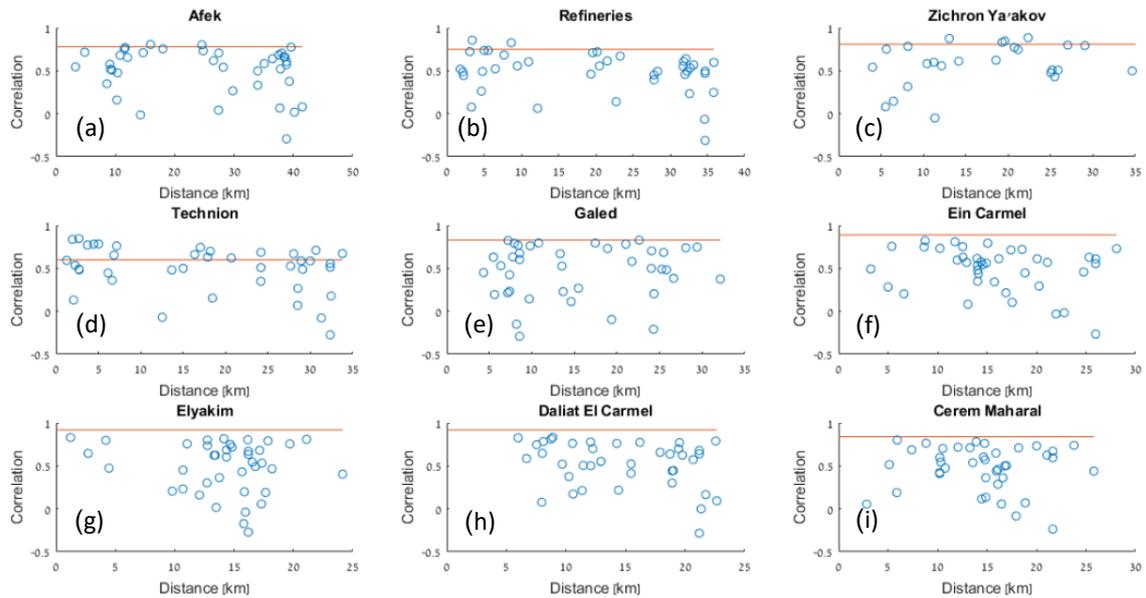

**Figure 8:** Correlation tests. Panels (a) to (i) present the test results as obtained at each humidity gauge location. The horizontal red line indicates the correlation value between an estimate based on all of the links and the measurements of the humidity gauges located across the test site. The correlation values of each separate link against the stations is marked by a circle.

In a similar fashion, Figure 9 presents the RMSD between humidity estimates derived from weighting all of the links at the points where humidity gauges were located compared to the correlation of the humidity measurements taken by each separate link from the link group, and the humidity gauge.

It can be seen that the RMSDs based on weighting all of the links are significantly lower (excepting a small number of cases) than the RMSDs between the measurements of each single link and the station measurements.

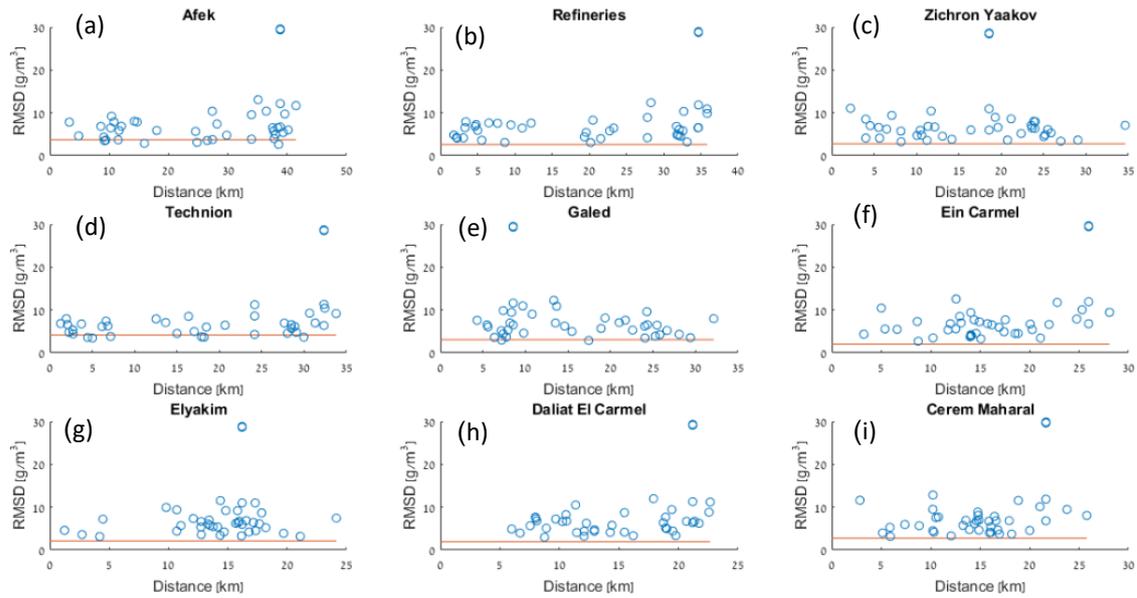

**Figure 9:** RMSD tests. The horizontal red line indicates the RMSD between an estimate based on all of the links and the measurements of the humidity gauges at the test site humidity gauge stations. The RMSD values of each separate link against the stations is marked by a blue circle.

## 5 SUMMARY AND CONCLUSIONS

Although the humidity field is a key atmospheric parameter, it is one of the least well understood variables due to the difficulty in measuring it using existing techniques. As a result, prediction of convective precipitation, on the storm scale, is limited (National Research Council, 1998, Fabry 2006).

Two earlier works (David et al., 2009, 2011) have shown the feasibility of estimating atmospheric humidity using a single commercial microwave link. The current study extends that prior research and demonstrates the potential of estimating the whole humidity field in space by combining measurements from a large number of different sources (MWLs) as a network of multi sensors. The results analysis (Fig. 8, Fig. 9) shows that estimating the humidity field using measurements from a large number of links significantly improves performance when compared to estimation using a single link. That being said, there are a small number of single links that provided lower RMSD/higher correlation with the humidity gauges even when compared to an estimate based on a large number of links. Future research can focus on characterizing specific links which have a higher humidity estimating potential when compared to the rest.

The theoretical simulation tool developed, indicates the system's potential to provide 2D humidity maps on a country wide scale. Additionally, the simulation points to the future capability of e band microwave systems that are being deployed these days, to provide higher resolution measurements when compared with links operating in the 6–40 GHz range. Thus, a near ground sensor network deployed widely around the world, may potentially allow, for the first time, observation of the humidity field in high resolution, and studying of this central atmospheric parameter in ways that were not previously possible due to the limitations of conventional instruments.

Commercial microwave link humidity measurements can suffer from environmental interference factors, such as – dew condensation on the antennas (David et al., 2016), rain and fog, which induce additional attenuation to the system (Rec. ITU-R P.838-3, 2004; Rec. ITU-R P.840-4, 2009), changes in atmospheric refractivity index that may cause bending of the microwave ray and variations in RSL as a result (Rec. ITU-R P.453-10, 2012; David and Gao, 2018) and more (Zinevich et al., 2010). Technical factors that affect measurement quality include the given magnitude resolution of the system (In this case 0.1 dB per link, however there are links with measurement errors of 1 dB for example), white noise and more.

Following the above discussion, one can note that the reason for signal attenuation may stem from various sources (both environmental and technical) which can often occur simultaneously such as dew and fog or rain and humidity variations. Thus, an additional future challenge is to test the ability to detect and separate the different sources of attenuation. Of course, one possibility is to perform this separation using specialized meteorological instruments, if available. For example, in this study we used rain gauge records to rule out the existence of rainfall across the test site. However, there may be cases in which attenuation source separation can be performed without the use of dedicated instruments. For example, the ability to automatically detect rain (Goldshtein et al., 2009; Chwala et al., 2012) or to separate rain from sleet (Cherkassky et al., 2014) using pure RSL measurements has been demonstrated. Another study indicated the potential to detect dew (Harel et al., 2015). Commercial MWLs operate at different frequencies and polarizations, spanning different lengths and deployed at various heights. Furthermore, the manner in which the measurements are acquired by the network is also varied. For example, some microwave systems provide mean RSL measurements or Max/Min RSL measurements per time interval.

Other networks provide instant RSL records etc. On the other hand, different weather conditions have a different effect on the microwave channel. Thus, for example, rain causes a decrease in signal which is typically significantly higher than that induced by humidity or fog. Dew causes wetting on the antennas themselves, and therefore MWLs of different path lengths are similarly affected by this phenomenon (in contrast to the case of rain, for instance, where the intensity of attenuation depends on the length of the link), water vapor causes higher attenuation around its absorption lines (David et al., 2009), etc. Thus, the combination of the system's diversity and the unique features of each weather phenomenon may allow, in some cases, for classification without any assistance or with minimum assistance from specialized equipment. Further investigation, which is left for future research, is required on this topic.

Over all, the results reported in this study point to the strong potential of the proposed technique to derive 2D humidity information. Never the less, transforming this concept into practical application requires additional examination and further research: during this study, link calibration was done using side information taken from humidity gauges prior to the start of measurements. Future research could focus on minimizing or obviating the dependence on additional information besides the measurements received from the links themselves. An example of such a possibility is the use of estimates of absolute humidity derived from numerical forecasting models for the initial calibration. Another possibility is to calculate the expected RSL theoretically without including the contribution to the attenuation due to atmospheric humidity and to examine the ability to receive moisture observations that are independent of any prior atmospheric information. Furthermore, in order to better understand the limitations of the technique and its performance, it should be observed in future research campaigns over additional time periods, in different regions of the world, and during different seasons.

Notably, these present and previous works (David et al., 2009; 2011) open the possibility of assimilation of microwave links signal attenuation or their retrieved moisture (and/or precipitation) into high-resolution meteorological forecasting models. Several aspects that are crucial for successful assimilation will have to be considered with careful scrutiny (see e.g. Kalnay, 2002). First, an estimate of the associated observation and representativeness error. A first step into this matter is shown in the present manuscript (e.g. comparison to gauges, Figs. 8 and 9). Second, present forecasting models are limited to ~ 1–2 km grid-size as finest horizontal resolution. This implies that even though MWLs information may be available at finer horizontal resolution it will have to be compatible with model resolution by creation of coarser "super-observations" and possibly introduce the sub-grid inhomogeneity in a parameterized mode. Finally, we note that all modern data assimilation techniques enable direct assimilation of the attenuation information provided a suitable model forward operator, thus, avoiding assumptions in the moisture retrieval process. Such a model forward operator can be derived following eq. (1). If the attenuation is significantly nonlinear, the forward operator may need to account for sub-grid variability in order to predict area/time averages accurately.

Overall, the results obtained point to the potential for improving the ability to cope with the dangers associated with extreme weather. Flash floods, for example, are typically triggered as a result of intense rainfall. The "fuel" for the formation of convective rain cells that lead to such rainfall is water vapor. Earlier warning, and more precise prediction means better ability to contend with dangers to humans and their environment.

**References**

AGU (1995), Water Vapor in the Climate System, special report, AGU, Washington, D. C.

Cherkassky, D., Ostrometzky, J., & Messer, H. (2014). Precipitation classification using measurements from commercial microwave links. *IEEE Transactions on Geoscience and Remote Sensing*, *52*(5), 2350-2356.

Chwala, C., Gmeiner, A., Qiu, W., Hipp, S., Nienaber, D., Siart, U., ... & Kunstmann, H. (2012). Precipitation observation using microwave backhaul links in the alpine and pre-alpine region of Southern Germany. *Hydrology and Earth System Sciences*, *16*(8), 2647-2661.

Chwala, C., Gmeiner, A., Qiu, W., Hipp, S., Nienaber, D., Siart, U., ... & Kunstmann, H. (2012). Precipitation observation using microwave backhaul links in the alpine and pre-alpine region of Southern Germany. *Hydrology and Earth System Sciences*, *16*(8), 2647-2661.

Chwala, C., Kunstmann, H., Hipp, S., & Siart, U. (2014). A monostatic microwave transmission experiment for line integrated precipitation and humidity remote sensing. *Atmospheric Research*, 144, 57-72.

Csurgai-Horváth, L., & Bitó, J. (2010, June). Fog attenuation on V band terrestrial radio and a low-cost measurement setup. In *Future Network and Mobile Summit, 2010* (pp. 1-9). IEEE.

David, N., Alpert, P., & Messer, H. (2009). Technical Note: Novel method for water vapour monitoring using wireless communication networks measurements. *Atmospheric Chemistry and Physics*, *9*(7), 2413-2418.

David, N., Messer, H., & Alpert, P. (2011). *Humidity measurements using commercial microwave links*. INTECH Open Access Publisher.

David, N., Alpert, P., & Messer, H. (2013a). The potential of cellular network infrastructures for sudden rainfall monitoring in dry climate regions. *Atmospheric Research*, *131*, 13-21.

David, N., Alpert, P., & Messer, H. (2013b). The potential of commercial microwave networks to monitor dense fog-feasibility study. *Journal of Geophysical Research: Atmospheres*, *118*(20).

David, N., Sendik, O., Messer, H., & Alpert, P. (2015). Cellular network infrastructure: the future of fog monitoring?. *Bulletin of the American Meteorological Society*, *96*(10), 1687-1698.

David, N., Harel, O., Alpert, P., & Messer, H. (2016, March). Study of attenuation due to wet antenna in microwave radio communication. In *Acoustics, Speech and Signal Processing (ICASSP), 2016 IEEE International Conference on* (pp. 4418-4422). IEEE.

David, N., & Gao, H. O. (2016). Using Cellular Communication Networks To Detect Air Pollution. *Environmental science & technology*, *50*(17), 9442-9451.

David, N., & Gao, H. O. (2018). Using Cell-Phone Tower Signals for Detecting the Precursors of Fog. *Journal of Geophysical Research: Atmospheres*, *123*(2), 1325-1338.

Ducrocq, V., Ricard, D., Lafore, J. P., & Orain, F. (2002). Storm-scale numerical rainfall prediction for five precipitating events over France: On the importance of the initial humidity field. *Weather and Forecasting*, *17*(6), 1236-1256.

Ericsson. (2015): Microwave towards 2020 report.

Fabry, F. (2006). The spatial variability of moisture in the boundary layer and its effect on convection initiation: Project-long characterization. *Monthly Weather Review*, *134*(1), 79-91.

Fencl, M., Rieckermann, J., Schleiss, M., Stránský, D., & Bareš, V. (2013). Assessing the potential of using telecommunication microwave links in urban drainage modelling. *Water Science and Technology*, *68*(8), 1810-1818.

Goldshtein, O., Messer, H., & Zinevich, A. (2009). Rain rate estimation using measurements from commercial telecommunications links. *Signal Processing, IEEE Transactions on*, *57*(4), 1616-1625.

Harel, O., David, N., Alpert, P., & Messer, H. (2015). The Potential of Microwave Communication Networks to Detect Dew—Experimental Study. *Selected Topics in Applied Earth Observations and Remote Sensing, IEEE Journal of*, *8*(9), 4396-4404.

Held, I. M., & Soden, B. J. (2000). Water vapor feedback and global warming 1. *Annual review of energy and the environment*, *25*(1), 441-475.

Hoedjes, J. C., Kooiman, A., Maathuis, B. H., Said, M. Y., Becht, R., Limo, A., ... & Su, B. (2014). A conceptual flash flood early warning system for Africa, based on terrestrial microwave links and flash flood guidance. *ISPRS International Journal of Geo-Information*, *3*(2), 584-598.

Kalnay, E (2012). Atmospheric Modeling, Data Assimilation and Predictability. Cambridge University Press, New York.

Rayitsfeld, A., Samuels, R., Zinevich, A., Hadar, U., & Alpert, P. (2012). Comparison of two methodologies for long term rainfall monitoring using a commercial microwave communication system. *Atmospheric research*, *104*, 119-127.

Rec. ITU-R P.838-3: Specific attenuation model for rain for use in prediction methods, 2005

Rec. ITU-R P.676-6: Attenuation by atmospheric gases, 2005.

Rec. ITU-R P.840-4: Attenuation due to clouds and fog, 2009.

Rec. ITU-R P.453-10: The radio refractive index: its formula and refractivity data, 2012

Leijnse, H., Uijlenhoet, R., & Stricker, J. N. M. (2007). Rainfall measurement using radio links from cellular communication networks. *Water Resources Research*, *43*(3).

Messer, H., Zinevich, A., & Alpert, P. (2006). Environmental monitoring by wireless communication networks. *Science*, *312*(5774), 713-713.

Morin, E., Jacoby, Y., Navon, S., & Bet-Halachmi, E. (2009). Towards flash-flood prediction in the dry Dead Sea region utilizing radar rainfall information. *Advances in water resources*, *32*(7), 1066-1076.

Müller, R., Kunz, A., Hurst, D. F., Rolf, C., Krämer, M., & Riese, M. (2015). The need for accurate long-term measurements of water vapor in the upper troposphere and lower stratosphere with global coverage. *Earth's Future*.

Overeem, A., Leijnse, H., & Uijlenhoet, R. (2013). Country-wide rainfall maps from cellular communication networks. *Proceedings of the National Academy of Sciences*, *110*(8), 2741-2745.

Schmidt, G. A., Ruedy, R. A., Miller, R. L., & Lacis, A. A. (2010). Attribution of the present-day total greenhouse effect. *Journal of Geophysical Research: Atmospheres*, *115*(D20).

Shepard, D. (1968, January). A two-dimensional interpolation function for irregularly-spaced data. In *Proceedings of the 1968 23rd ACM national conference* (pp. 517-524). ACM.

Sherwood, S. C., Roca, R., Weckwerth, T. M., & Andronova, N. G. (2010). Tropospheric water vapor, convection, and climate. *Reviews of Geophysics*, *48*(2).

Trenberth, K. E. (1999). Conceptual framework for changes of extremes of the hydrological cycle with climate change. In *Weather and Climate Extremes* (pp. 327-339). Springer Netherlands.

Uijlenhoet, R., Overeem, A., & Leijnse, H. (2018). Opportunistic remote sensing of rainfall using microwave links from cellular communication networks. *Wiley Interdisciplinary Reviews: Water*.

Van Vleck, J. H. (1947). The absorption of microwaves by uncondensed water vapor. *Physical Review*, *71*(7), 425.

Weckwerth, T. M. (2000). The effect of small-scale moisture variability on thunderstorm initiation. *Monthly weather review*, *128*(12), 4017-4030.

Weckwerth, T. M., Murphey, H. V., Flamant, C., Goldstein, J., & Pettet, C. R. (2008). An observational study of convection initiation on 12 June 2002 during IHOP_2002. *Monthly Weather Review*, *136*(7), 2283-2304.

World Meteorological Organization (WMO): Guide to meteorological instruments and methods of observation, 7th edition, ISBN 978-92-63-10008-5, 2008.

Zinevich, A., Messer, H., & Alpert, P. (2009). Frontal rainfall observation by a commercial microwave communication network. *Journal of Applied Meteorology and Climatology*, *48*(7), 1317-1334.

Zinevich, A., Messer, H., & Alpert, P. (2010). Prediction of rainfall intensity measurement errors using commercial microwave communication links. *Atmospheric Measurement Techniques*, *3*(5), 1385-1402.

**Acknowledgements**

We wish to sincerely thank Cellcom, Pelephone, and PHI who provided the microwave data for our research. In Cellcom, we would like to thank: E. Levi, Y. Koriat, B. Bar and I. Alexandrovitz. In Pelephone, we would like to thank: N. Dvela, A. Hival and Y. Shachar. In PHI we would like to thank : Y. Bar Asher, O. Tzur, Y.

Sebton, A. Polikar and O. Borukhov. We deeply thank the Israeli Meteorological Service (IMS) and the Israeli Ministry of Environmental Protection for providing meteorological data.

This work was supported by a grant from the German Research Foundation (DFG) through the project "Integrating Microwave Link Data For Analysis of Precipitation in Complex Terrain: Theoretical Aspects and Hydrometeorological Applications" (IMAP).Authors Noam David and H. Oliver Gao acknowledge partial funding support by National Science Foundation project CMMI-1462289, the Natural Science Foundation of China (NSFC) project # [71428001](#), Cornell Atkinson Center for a sustainable future, and the Lloyd's Register Foundation, UK.